\documentclass{emulateapj}

\slugcomment{ApJ Letters, accepted}

\shorttitle{Accretion rates in T Tauri stars and brown dwarfs}
\shortauthors{Alexander \& Armitage}

\begin{document}

\title{The stellar mass-accretion rate relation in T Tauri stars and brown dwarfs}

\author{Richard~D.~Alexander\altaffilmark{1} and Philip~J.~Armitage\altaffilmark{1,2}}
\altaffiltext{1}{JILA, 440 UCB, University of Colorado, Boulder, CO 80309-0440}
\altaffiltext{2}{Department of Astrophysical and Planetary Sciences, University of Colorado, Boulder, CO 80309-0391}
\email{rda@jilau1.colorado.edu}

%%%%%%%%%%%%%%%%%%%%%%%

\begin{abstract} 
Recent observations show a strong correlation between stellar mass and accretion rate in young stellar and sub-stellar objects, with the scaling $\dot{M}_{acc} \propto M_*^2$ holding over more than four orders of magnitude in accretion rate. We explore the consequences of this correlation in the context of disk evolution models.  We note that such a correlation is not expected to arise from variations in disk angular momentum transport efficiency with stellar mass, and suggest that it may reflect a systematic trend in disk initial conditions. In this case we find that brown dwarf disks initially have rather larger radii than those around more massive objects.  By considering disk evolution, and invoking a simple parametrization for a shut-off in accretion at the end of the disk lifetime, we show that such models predict that the scatter in the stellar mass-accretion rate relationship should increase with increasing stellar mass, in rough agreement with current observations. 
\end{abstract}


%%%%%%%%%%%%%%%%%%%%%%

\section{Introduction} It is well established that young stars of around solar mass, such as T Tauri stars (henceforth TTs), are surrounded by disks. Reprocessing of stellar radiation by these disks causes excess emission above the stellar photosphere in the infrared, and the accretion of material on to the stellar surface produces both line emission and excess blue continuum emission. This permits an observational measurement of the instantaneous stellar accretion rate for large samples of TTs \citep[e.g][]{gullbring98}. More recently, it has become clear that young brown dwarfs (henceforth BDs) also posses circumstellar disks, with observations detecting infrared excesses \citep[e.g.][]{nt01,rayjay03b}, accretion signatures \citep*[e.g.][]{rayjay03a,natta04,subu05} and millimeter continuum emission \citep[e.g.][]{klein03}.

If BD formation is merely a scaled-down version of star formation, one might guess that the disk mass and stellar accretion rate $\dot{M}_{acc}$ should scale roughly proportional to the (sub-)stellar mass $M_*$. In fact, the accretion rate on to the ``stellar'' surface shows a striking correlation with the {\em square} of the stellar mass \citep{muz03,natta04,calvet04,muz05,subu05}.  The correlation shows a large scatter, but holds over more than 2 orders of magnitude in mass and 4 orders of magnitude in accretion rate. We expect a large scatter, as disk accretion rates decrease as disks evolve \citep[e.g.][]{hcga98}, but the physical origin of the $\dot{M}_{acc} \propto M_*^2$ relationship is not clear. Here we suggest that the most conservative interpretation is that the correlation reflects the initial conditions established when the disk formed, followed by subsequent viscous evolution of the disk. We explore the observational consequences of this model.

%%%%%%%%%%%%%%%%%%%%%%%%%%%%%%%%%%%%%%%%%%%%%

\section{Model}\label{sec:model}
The equation for the surface density evolution, $\Sigma(R,t)$, of a viscous accretion disk is
\begin{equation}\label{eq:1ddiff}
\frac{\partial \Sigma}{\partial t} = \frac{3}{R}\frac{\partial}{\partial R}\left[ R^{1/2} \frac{\partial}{\partial R}\left(\nu \Sigma R^{1/2}\right) \right]  \, ,
\end{equation}
where $R$ is cylindrical radius, $t$ is time, and $\nu(R,t)$ is the kinematic viscosity of the disk \citep{pringle81}.  If the viscosity is independent of time and can be expressed as a power-law function of radius $\nu \propto R^{\gamma}$, then Equation \ref{eq:1ddiff} permits exact analytic solutions.  Here we consider the similarity solution of \citet{hcga98}, after \citet{lbp74}.  In this case the solution for the surface density of the evolving disk is
\begin{equation}
\Sigma(R,t) = \frac{M_d(0) (2-\gamma)}{2\pi R_0^2 r^{\gamma}} \tau^{\frac{-(5/2 - \gamma)}{2-\gamma}} \exp\left(-\frac{r^{2-\gamma}}{\tau}\right) \, ,
\end{equation}
where $M_d(0)$ is the initial disk mass.  The dimensionless radius $r=R/R_0$, where $R_0$ is a scale radius which sets the initial disk size.  The dimensionless time $\tau=t/t_{\nu} + 1$, where the viscous scaling time $t_{\nu}$ is given by
\begin{equation}\label{eq:tnu}
t_{\nu} = \frac{R_0^2}{3(2-\gamma)^2\nu_0} \, .
\end{equation}
Here $\nu_0$ is the value of the viscosity at radius $R_0$.  The accretion rate on to the star is therefore
\begin{equation}\label{eq:mdot_t}
\dot{M}_{acc} = \frac{M_d(0)}{2(2-\gamma)t_{\nu}} \tau^{\frac{-(5/2 - \gamma)}{2-\gamma}} \, .
\end{equation}
We suggest that the time evolution implied by this equation is responsible for some of the observed scatter in the mass-accretion rate distribution, and assume that the observed $\dot{M}_{acc} \propto M_*^2$ relationship holds for the initial accretion rates.  Observations of BD disks have shown that the disk-to-star mass ratio is comparable to that found for TTs \citep{klein03}, so we assume that the initial disk mass scales linearly with the stellar mass.  In order to reproduce the observed correlation, that $\dot{M}_{acc} \propto M_*^2$, we therefore require that the
\begin{equation}\label{eq:tnu_mdot}
t_{\nu}\propto \frac{1}{M_*} \, .
\end{equation}
For TTs, the viscous time-scale $t_{\nu}$ is typically $\sim 10^4$--$10^5$yr \citep[e.g.][]{hcga98}, and therefore we expect that the viscous scaling time for BDs be of order $10^6$yr.

However we note that while current observations \citep[e.g.][]{klein03} suggest a linear relationship between disk and stellar mass the data do not yet rule out a steeper relationship.  A much steeper scaling of $M_d \propto M_*^2$ could also explain the observed  $\dot{M}_{acc} \propto M_*^2$ relationship without requiring the change in $t_{\nu}$ with stellar mass seen in Equation \ref{eq:tnu_mdot}.  However in this case BD disks are less massive (relative to the central object) then their TT counterparts, with few BDs having disks more massive than $\sim 1$M$_J$.  This seems unlikely, but current observations do not yet rule it out.

We note that it is also possible to satisfy the observed relationship if $t_{\nu}$ {\em increases} with increasing stellar mass, so that $\dot{M}_{acc} \propto M_*^2$ only at late times (i.e.~$t \gg t_{\nu}$).  We reject this scenario for three reasons.  Firstly, it requires extremely short viscous time-scales for BD disks ($t_{\nu} < 10^3$yr), which seem unlikely.  Also, this scenario requires that the initial accretion rate be higher for BDs than TTs, which again seems unlikely.  Lastly, in this case most of the initial disc mass is accreted very rapidly. By $t\sim 1$Myr BD disk masses are only a few percent of a Jupiter mass, which is rather less massive than observed \citep{klein03}.

\subsection{Viscosity}
We now consider the consequences of this relationship in terms of a simple alpha-model for the disk viscosity \citep{ss73}.  Theoretical studies of angular momentum transport in disks suggest that magnetohydrodynamic (MHD) turbulence can produce values of $\alpha$ of order 0.01 \citep{brandenburg95,stone96,balbus98}.  In a Keplerian disk we can re-express this form for the viscosity in terms of the disk midplane temperature $T$ and the angular frequency $\Omega$
\begin{equation}
\nu \propto \alpha  \frac{T}{\Omega} \propto \alpha \frac{T R^{3/2}}{M_*^{1/2}} \, ,
\end{equation}
If we substitute the viscosity at the scale radius $R_0$ into Equation \ref{eq:tnu} we find that the viscous scaling time of the disk satisfies
\begin{equation}
t_{\nu} \propto \frac{R_0^{1/2} M_*^{1/2}}{\alpha T(R_0)} \, .
\end{equation}
MHD simulations suggest that $\alpha$ does not vary significantly with $M_*$.  Also, constant aspect ratio $H/R$ demands that the disk midplane temperature scales as $T \propto M_*/R$.  Therefore in the case of constant aspect ratio $H/R$ we can solve for the dependence of $R_0$ with $M_*$ exactly to find $R_0 \propto M_*^{-1/3}$.  In real TT and BD disks, however, reprocessing of stellar radiation is the dominant mode of disk heating and typically results in a "flaring" disk structure \citep[e.g.][]{kh87,cg97}.  In this case the temperature scales as $T \propto R^{-1/2}$, which in turn gives $R_0 \propto M_*^{-1/2}$.  The exact form of the disk midplane temperature is therefore somewhat model-dependent, but it is clear that in order to satisfy Equation \ref{eq:tnu_mdot} we require that $R_0$ decrease with increasing stellar mass.  This implies that BD disks are initially larger than those around TTs, with typical scale radii $R_0$ for BD disks of around 50--100AU.  This is not altogether unexpected, as if the material in a collapsing cloud has a roughly constant specific angular momentum we expect smaller disks around more massive objects.  We note that the disks expand in size on their viscous time-scale, and therefore TTs disks will rapidly expand to be comparable in size to BD disks.  Additionally, external factors, such as tidal truncation by a binary companion \citep{pac77}, can strongly influence the size of an evolving disk.  It is therefore unlikely that this predicted scaling of the initial disk size with stellar mass will have a significant observational signature.  

\subsection{Cessation of accretion}
Observations of TTs show that the the cessation of accretion occurs concurrently with (inner) disk clearing \citep[e.g.][]{hartigan90}, and the data are consistent with individual objects undergoing a rapid ``shut-off'' in accretion once the accretion rate falls to some low level \citep*{acp03}.  It has been suggested that photoevaporation by the central star is responsible for this shut-off in accretion \citep*{cc01}, with accretion terminating once the disk accretion rate falls to a level comparable to the expected photoevaporative disk wind rate.  The integrated mass-loss rate from a photoevaporative disk wind is given by
\begin{equation}\label{eq:mdot_photoevap}
\dot{M}_w = 1.6\times 10^{-10}\mathrm M_{\odot} \mathrm{yr}^{-1} \left(\frac{\Phi}{10^{41}\mathrm s^{-1}}\right)^{1/2} \left(\frac{M_*}{1\mathrm M_{\odot}}\right)^{1/2} \, ,
\end{equation}
where $\Phi$ is the rate at which the star emits ionizing photons \citep{holl94,font04}.  \citet*{chrom} show that chromospheric activity in TTs can provide an ionizing flux of order $10^{41}$photons s$^{-1}$ or higher.  The photospheric emission from both TTs and BDs is powered by energy released from gravitational collapse, but little is known about the magnetic activity of BDs.  However recent observations have found that the ratio of the X-ray:bolometric luminosity is approximately the same for BDs as for TTs \citep{coup_bd}, so we make the crude assumption that $\Phi$ scales linearly with mass.  Thus the predicted photoevaporative wind rate for a 0.01M$_{\odot}$ BD is of order $10^{-12}$M$_{\odot}$yr$^{-1}$.

\subsection{Model parameters}

\begin{figure}
\includegraphics[angle=270,width=\hsize]{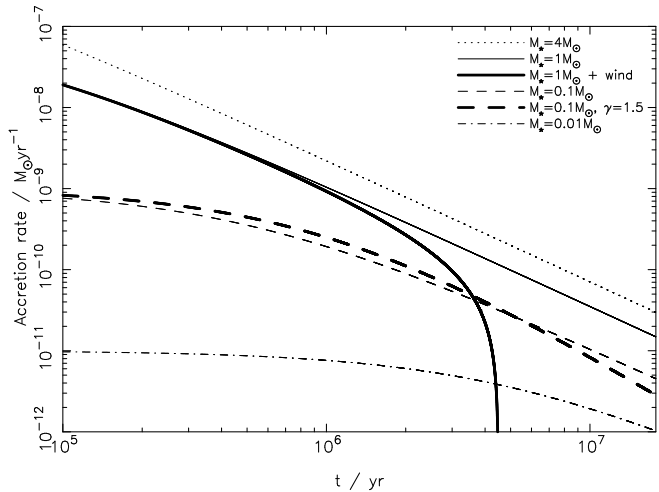}
\caption{Evolution of accretion rate in viscous disk models with varying stellar mass.  The models have initial disk masses of 0.01$M_*$ and viscosity index $\gamma=1$, and the viscous scaling time for the fiducial 1M$_{\odot}$ model is $5\times 10^4$yr.  The heavy solid line shows how the accretion rate in the fiducial model is shut-off by a photoevaporative wind.  The heavy dashed line shows how the evolution of the accretion rate changes if a different viscosity index ($\gamma=3/2$) is adopted (with the same initial accretion rate).}\label{fig:mdot_t}
\end{figure}

We have constructed disk evolution models in which the initial disk mass is given by $M_d(0) = 0.01M_*$, and the viscous scaling time of the disk is given by
\begin{equation}
t_{\nu} = 5 \times 10^4 \mathrm {yr} \left(\frac{M_*}{1\mathrm M_{\odot}}\right)^{-1} \, .
\end{equation}
We note, however, that some scatter in both $M_d(0)$ and $t_{\nu}$ is likely.  In our fiducial model we adopt a viscosity index $\gamma=1$, and we have evaluated models across the mass range $M_* = 0.01$--10M$_{\odot}$.  $\gamma=1$ is consistent with a $T \propto R^{-1/2}$ flared disc model of the type discussed above \citep[e.g.][]{kh87,cg97}, but $\gamma$ is not especially well-constrained by current observations \citep[see e.g.~discussion in][]{hcga98}.  Fig.\ref{fig:mdot_t} shows how the accretion rate in the models evolves with time for different stellar masses and viscosity laws: we see that the accretion history of these models is not strongly dependent on the choice of $\gamma$ over the timescales in which we are interested.  Also shown in Fig.\ref{fig:mdot_t} is the evolution of a model which incorporates disk photoevaporation \citep*[using the models of][]{alexander05}.  In this case the accretion rate is unchanged over most of the evolution, but accretion is rapidly shut-off at the end of the disk lifetime.  

\begin{figure}
\includegraphics[angle=270,width=\hsize]{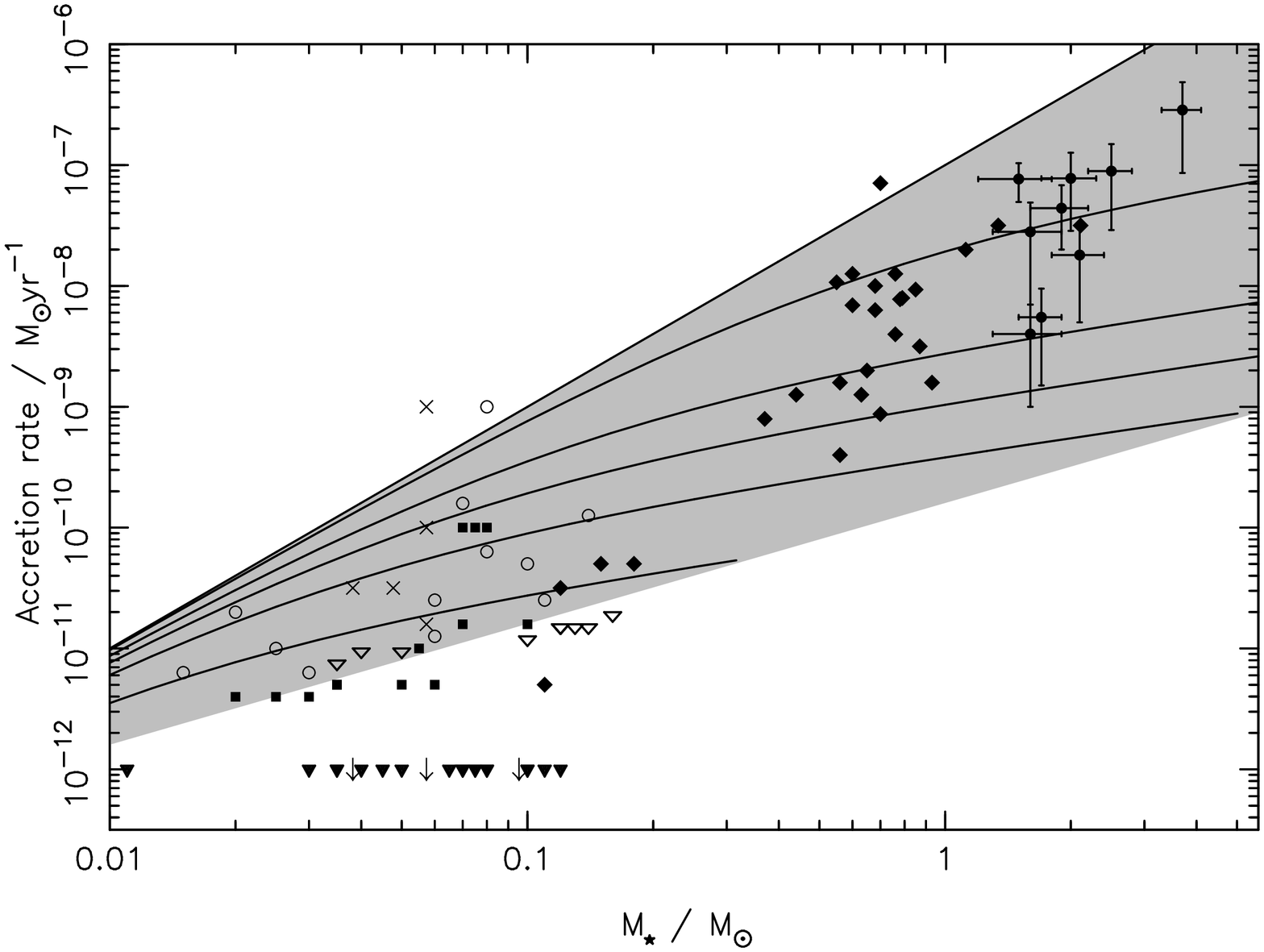}
\caption{Accretion rate plotted as a function of stellar mass.  Data points are taken from \citet[][filled circles with error bars]{calvet04}, \citet[][crosses for detections, arrows for upper limits]{natta04}, \citet[][filled squares for detections, filled triangles for upper limits]{muz05}, and \citet[][open circles for detections, open triangles for upper limits]{subu05}.  The diamonds show the compilation of previous data given in Table 4 of \citet{subu05}.  The solid lines are isochrones for our disk model, plotted (from top to bottom) at $t=0$, 0.1, 0.5, 1.0, 2.0 and 5.0Myr.  The shaded area indicates the range of predicted accretion rates when accretion is shut-off by photoevaporation at late times.}\label{fig:mdot_mstar}
\end{figure}

We have evaluated the above model set but assumed that accretion is shut-off once the accretion rate falls below some critical value.  We define this value as in Equation \ref{eq:mdot_photoevap}, and assume that the ionizing flux scales as $\Phi = 10^{41}\mathrm s^{-1} \left(M_*/1\mathrm M_{\odot}\right)$.  Fig.\ref{fig:mdot_mstar} shows how accretion rate varies as a function of stellar mass in these models, and also how the accretion rates change as the disks evolve.  For comparison we include observed data points from a number of recent studies.  Our simple model reproduces both the slope and scatter in the observed distribution well, from the BD regime up to masses of several M$_{\odot}$.

The assumptions in our model lead to some notable observable consequences.  The longer viscous scaling time for disks around lower mass objects means that the evolution of these disks is slower (see Fig.\ref{fig:mdot_t}).  However the assumed scaling of $\Phi$ with stellar mass means the rate at which accretion is shut-off decreases only linearly with stellar mass, and so BD disks evolve for fewer viscous times than TT disks before accretion ceases (see Fig.\ref{fig:mdot_mstar}).  As a result we expect a smaller spread in observed accretion rates as stellar mass decreases, with BD accretion rates showing less scatter than those for TTs.  This is in tentative agreement with the data shown in Fig.\ref{fig:mdot_mstar}, where the detected accretion rates show a (2-$\sigma$) scatter of $\pm1.0$dex in the mass range 0.02--0.2M$_{\odot}$, but a larger scatter of $\pm1.4$dex in the range 0.4--4.0M$_{\odot}$.  The model also predicts that few, if any, objects should be detected with accretion rates significantly below $10^{-12}$M$_{\odot}$.  However we note that this value is near to the sensitivity limit of current observations, as seen by the locus of the many non-detections (upper limits) in Fig.\ref{fig:mdot_mstar}.  In this scenario we expect these objects to have little or no ongoing accretion, essentially appearing as ``weak-lined'' BDs.  Lastly, in our model both TT and BD disks have typical lifetimes of order a few Myr, with no significant difference in disk lifetimes across the mass spectrum.  Current observations tentatively agree with this prediction \citep{subu05}, but more data are needed in order to make a more detailed comparison.

%%%%%%%%%%%%%%%%%%%%%%%%%%%%%

\section{Discussion and Summary}
In the preceding Section we have outlined the observational consequences if the $M_*$-$\dot{M}_{acc}$ relation for T Tauri stars and brown dwarfs is set by disk initial conditions followed by viscous evolution.  Alternatively, the observed relation might reflect systematic variations in the structure or feeding of disks. One possibility is that some aspect of disk angular momentum transport introduces a bottleneck into the accretion flow, restricting the stellar accretion rate to the observed $\dot{M}_{acc} \propto M_*^2$ independent of anything happening at larger radii. \citet{subu05} suggest that this could be caused by suppression of the magnetorotational instability (MRI) due to a reduced ionization fraction in the disks around cooler, lower-mass stars, a scenario is similar to the layered disk model introduced by \citet{gammie96}.  Except for very close to the star, accretion occurs via an externally ionized surface layer of column $\Sigma_a$, at a rate $\dot{M}_{acc} \propto \alpha^2 \Sigma_a^3$. In the simplest versions of such models the dominant ionization source for the surface layer is cosmic rays, with the consequence that both $\Sigma_a \sim 10^2 \ {\rm gcm^{-2}}$, and the resulting stellar $\dot{M}_{acc}$, are constants {\em independent} of stellar mass. A variant of such a model might yield the observed $M_*$-$\dot{M}_{acc}$ relation, but it does not seem promising. 

A more radical possibility is that the disk acts only as a short-term buffer between the star and gas continually captured from the surrounding environment via Bondi-Hoyle accretion \citep{bondi44}. This immediately predicts the observed $\dot{M}_{acc} \propto M_*^2$ scaling \citep{padoan05}. For such a model to work, the captured gas must circularize at a small enough radius that the viscous timescale is short compared to the $\sim$Myr disk lifetime, and we would expect the accretion rate to vary with the circumstellar environment. However \citet{subu05} claim that many of the objects for which this correlation is seen to hold show similar accretion rates despite a broad dispersion in their environments.  Also, for Bondi-Hoyle accretion in a turbulent or clumpy medium, we expect the scatter in accretion rate to {\em increase} with decreasing stellar mass, as the smaller accretion radius of lower mass stars samples a smaller volume in which random fluctuations in density would be greater.

In this {\em Letter}, we have examined the consequences that follow if the observed $\dot{M}_{acc} \propto M_*^2$ relation for brown dwarfs and low mass stars reflects a combination of varying disk initial conditions with stellar mass and viscous disk evolution. A relation of this form requires that the ``initial'' disk size increase with decreasing stellar mass.  A larger initial disk size for brown dwarfs is not unreasonable if the specific angular momentum of the last material to be accreted is independent of stellar mass, though the specific form of the required scaling does not follow from any such simple consideration.  Consequently we predict that disks around isolated BDs should be rather large, typically 50--100AU in radius.  Future observations of disk sizes using millimeter interferometry should prove invaluable in testing this prediction.  

If, as we advocate, the observed $\dot{M}_{acc} \propto M_*^2$ scaling is the result of disk initial conditions, we have shown that both the scatter in the disk accretion rate for the population as a whole, and the change in mean accretion rate with stellar age, ought to be smaller for brown dwarfs than for more massive pre-main-sequence stars (such as TTs). This provides an observational test for distinguishing between this model and alternatives in which the relation arises from either disk physics \citep{subu05} or Bondi-Hoyle accretion \citep{padoan05}.

Lastly, following on from Section \ref{sec:model}, we note that current data do not yet rule out a steeper $M_*$-$M_d$ relationship than the linear scaling we have assumed.  A much steeper scaling of $M_d \propto M_*^2$ could also explain the observed accretion rate-stellar mass relationship without requiring the (spatially) large BD disks predicted by our model.  However in this case BD disks should have rather low masses, with few BD disks more massive than $\sim 1$M$_J$.  Thus in order to for viscous disk accretion to reproduce the observed $\dot{M}_{acc} \propto M_*^2$ relationship BD disks must either be spatially large, as we suggest, or have low masses.  Future observations of the sizes and masses of BD disks will be invaluable in resolving this degeneracy.

%%%%%%%%%%%%%%%%%%%%%%%%%%%%

\acknowledgments
We thank Cathie Clarke, Ray Jayawardhana and Giuseppe Lodato for interesting discussions, and an anonymous referee for useful comments.  This work was supported by NASA grants NAG5-13207, NNG04GL01G and NNG05GI92G, and by NSF grant AST~0407040.

%%%%%%%%%%%%%%%%%%%%%%%%%%

\end{document}